\documentclass[aps,prl,superscriptaddress,twocolumn,longbibliography,10pt]{revtex4-1}
\usepackage[bookmarks=true,colorlinks,citecolor=blue,urlcolor=blue]{hyperref}
\usepackage{ucs}
\usepackage[latin1]{inputenc}
\usepackage[english]{babel}
\usepackage[usenames, dvipsnames]{xcolor}
\usepackage[export]{adjustbox}
\usepackage{amsmath, 
			amsfonts,
			booktabs,
            microtype,
            mathtools, 
            graphicx,
            color,
            soul,
            nicefrac,
            float,
            url, 
            placeins,	
			braket,
			appendix} 	

\DeclareRobustCommand{\mb}[1]{%
  \ifmmode\text{\hl{$#1$}}\else\hl{#1}\fi
}

\begin{document}
\title{Realization of a Rydberg-dressed Ramsey interferometer and electrometer}
\author{A. Arias}
\affiliation{Physikalisches Institut, Universit\"at Heidelberg, Im Neuenheimer Feld 226, 69120 Heidelberg, Germany}\affiliation{IPCMS (UMR 7504) and ISIS (UMR 7006), University of Strasbourg and CNRS, 67000 Strasbourg, France}
\author{G. Lochead}\affiliation{Physikalisches Institut, Universit\"at Heidelberg, Im Neuenheimer Feld 226, 69120 Heidelberg, Germany}\affiliation{IPCMS (UMR 7504) and ISIS (UMR 7006), University of Strasbourg and CNRS, 67000 Strasbourg, France}
\author{T. M. Wintermantel}\affiliation{Physikalisches Institut, Universit\"at Heidelberg, Im Neuenheimer Feld 226, 69120 Heidelberg, Germany}\affiliation{IPCMS (UMR 7504) and ISIS (UMR 7006), University of Strasbourg and CNRS, 67000 Strasbourg, France}
\author{S. Helmrich}
\affiliation{Physikalisches Institut, Universit\"at Heidelberg, Im Neuenheimer Feld 226, 69120 Heidelberg, Germany}
\author{S. Whitlock}\email[e-mail: ]{whitlock@ipcms.unistra.fr}
\affiliation{Physikalisches Institut, Universit\"at Heidelberg, Im Neuenheimer Feld 226, 69120 Heidelberg, Germany}\affiliation{IPCMS (UMR 7504) and ISIS (UMR 7006), University of Strasbourg and CNRS, 67000 Strasbourg, France}
\date{\today}
	
\begin{abstract}
We present the experimental realization and characterization of a Ramsey interferometer based on optically trapped ultracold potassium atoms, where one state is continuously coupled by an off-resonant laser field to a highly-excited Rydberg state. We show that the observed interference signals can be used to precisely measure the Rydberg atom-light coupling strength as well as the population and coherence decay rates of the Rydberg-dressed states with sub-kilohertz accuracy and for Rydberg state fractions as small as one part in $10^{6}$. We also demonstrate an application for measuring small, static electric fields with high sensitivity. This provides the means to combine the outstanding coherence properties of Ramsey interferometers based on atomic ground states with a controllable coupling to strongly interacting states, thus expanding the number of systems suitable for metrological applications and many-body physics studies.
\end{abstract}
	
\maketitle

Ramsey interferometers involving trapped ensembles of neutral atoms or single ions have enabled the most precise measurements ever made. Besides their importance for defining time and frequency standards~\cite{gill2011,riehle2015}, they also hold great promise for searches for physics beyond the standard model~\cite{fisher2004newlimits,peik2004limit,derevianko2014,Huntemann2014,Godun2014}, exploring the physics of complex quantum systems~\cite{Martin2013}, and for realizing sensors capable of operating close to the Heisenberg limit~\cite{Jones2009,hosten2016measurement,facon2016sensitive}. By design however, the most precise Ramsey interferometers typically involve the coherent evolution of atoms that interact very weakly, either with one another or with external fields, seemingly precluding many possible applications. 

Here we demonstrate a Ramsey interferometer involving two magnetically insensitive hyperfine clock states, where one state is continuously coupled to a Rydberg state by an off-resonant laser field. This Rydberg-dressing approach provides the means to combine the outstanding coherence properties of atomic ground states with greatly enhanced sensitivity to external fields or controllable interparticle interactions mediated by the Rydberg state admixture. We show that strong Rydberg atom-light coupling can be reached with Ramsey coherence times that are orders of magnitude longer than the bare Rydberg state lifetime. We precisely measure the Rydberg-atom light coupling strength and independently determine the effective population decay and dephasing rates for the dressed-states, thereby identifying the dominant decoherence effects. Finally we demonstrate the suitability of the Rydberg-dressed Ramsey interferometer for metrological applications by precisely measuring a small applied electric field.
\begin{figure}[!ht]
	\centering	
	\includegraphics[width=1\columnwidth]{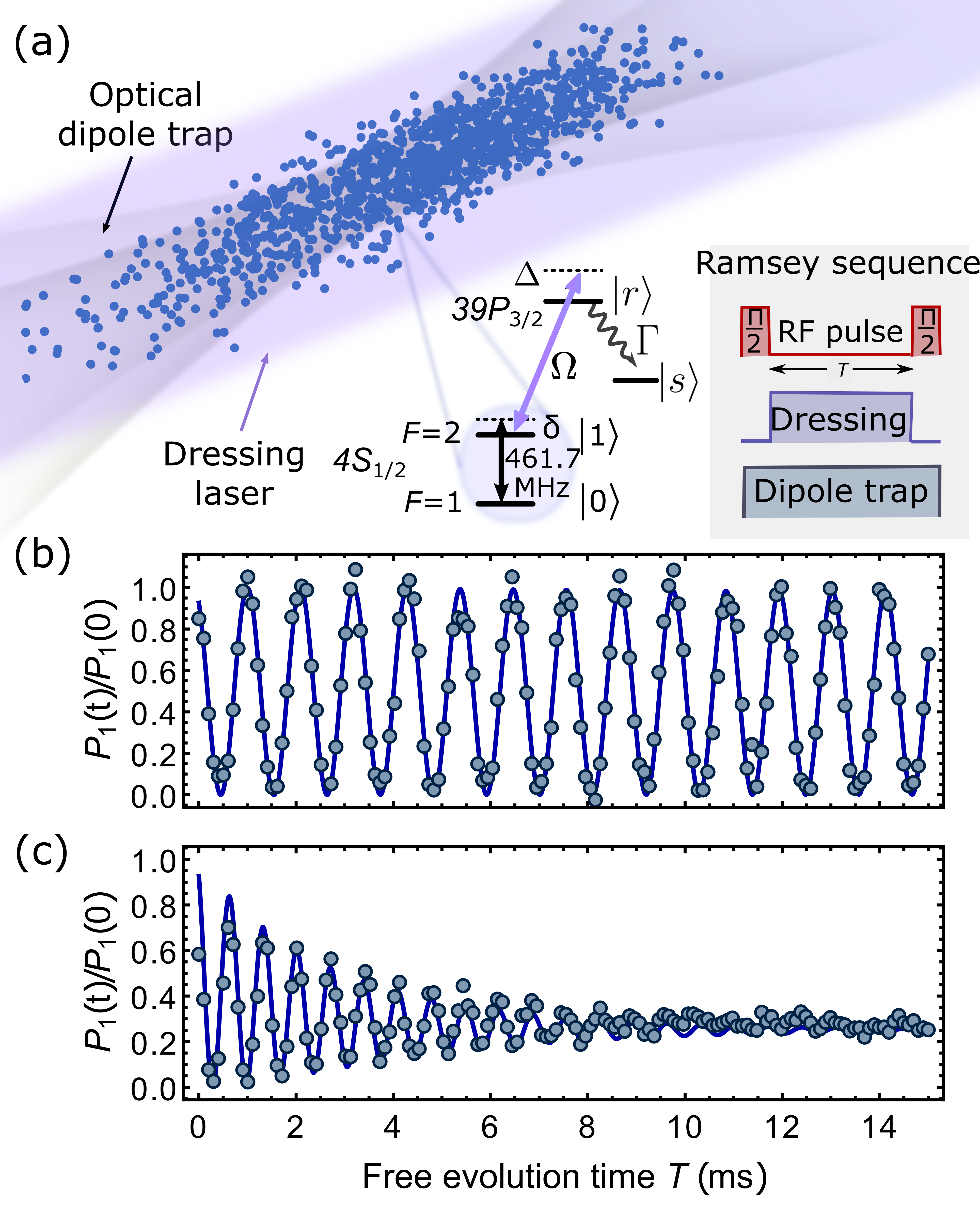}
	\caption{Experimental realization of a Rydberg-dressed Ramsey interferometer. (a) A cloud of $^{39}$K atoms is prepared in an optical dipole trap and uniformly illuminated by the Rydberg dressing laser. Each atom can be represented by a four level system involving the clock states $\ket{0}$ and $\ket{1}$, the $\ket{r=39P_{3/2}}$ Rydberg state and an auxiliary shelving $\ket{s}$. The $\ket{1}\rightarrow\ket{r}$ transition is coupled via the dressing laser with Rabi frequency $\Omega$ and detuning $\Delta$ during the Ramsey free evolution time $T$. (b) Ramsey fringes measured in the absence of the dressing laser showing very good coherence and (c) Ramsey fringes with the Rydberg excitation laser on with $\Delta/2\pi=-12.0$ MHz, showing a modified oscillation frequency and asymmetric decay of the interference contrast. The blue solid lines correspond to fits to the model described in the text.}
	\label{fig:ExperimentandRamsey}
\end{figure}

%

%
The starting point for our experiments is an ultracold gas of $^{39}$K atoms prepared in a crossed-beam optical dipole trap (shown in Fig.~\ref{fig:ExperimentandRamsey}a) following Refs.~\cite{salomon2014gray,arias2017versatile}. To realize the Ramsey interferometer we use the magnetically insensitive hyperfine ground states $\ket{0}=\ket{4S_{1/2}, F=1, m_F=0}$ and $\ket{1}=\ket{4S_{1/2}, F=2, m_F=0}$. The sample is spin-polarized in the initial $\ket{0}$ state using a sequence of optical pumping pulses and radio-frequency transfer pulses (similar to Ref.~\cite{Antoni-Micollier2017}). The final result is a sample of approximately $10^5$ atoms in the $\ket{0}$ state with a peak density of $1\times 10^{10}\,\mathrm{cm}^{-3}$ and temperature of $10\,\mu$K. From there we apply a Ramsey interferometry sequence consisting of two $\pi/2$ radio-frequency (RF) pulses ($90\,\mathrm{\mu s}$ each) separated by a variable evolution time $T$ (Fig.~\ref{fig:ExperimentandRamsey}a). The RF field originates from a programmable direct-digital-synthesis device which is referenced to a commercial rubidium standard. To read out the interferometer we measure the population in the $\ket{1}$ state using absorption imaging with a weak probe laser resonant to the $\ket{4S_{1/2}, F=2}\rightarrow \ket{4P_{3/2}, F=3}$ transition. 

Figure~\ref{fig:ExperimentandRamsey}b shows the measured Ramsey fringe pattern taken without the dressing field for free evolution times up to $15\,\mathrm{ms}$. The signal is very well described by a sinusoidal oscillation with no visible loss of contrast within this time. From the oscillation frequency we extract the Ramsey detuning of $\delta/2\pi=914.0(6)\,\mathrm{Hz}$. This corresponds to an absolute frequency for the $\ket{0}\rightarrow\ket{1}$ transition of $461.719\,748\,0(6)$\,MHz (statistical uncertainty in the fractional frequency of $\Delta f/f=1.3\times 10^{-9}$) which is within $30$\,Hz of previous measurements of the $^{39}\mathrm{K}$ hyperfine splitting~\cite{arimondo1977,Antoni-Micollier2017}. The small discrepancy can be accounted for by the expected quadratic Zeeman shift of $31(11)$\,Hz for bias field of $60(10)\,\mathrm{mG}$ and additional smaller corrections due to the differential AC Stark shift from the optical dipole trap and cold collisions.

Next we realize the Rydberg-dressed Ramsey interferometer by applying a laser field during the free evolution time, detuned by $\Delta/2\pi=-12.0\,\mathrm{MHz}$ from the $\ket{1}\rightarrow \ket{r=39P_{3/2}}$ transition. In the dressed-state picture~\cite{Dalibard1985}, this admixes a small Rydberg state component into the clock state (i.e. $\ket{\tilde 1}\approx \ket{1}+\sqrt{f_r}\ket{r}$, where $f_r$ is the relative population of the Rydberg state). The laser light is generated by frequency doubling a continuous-wave dye laser, with a second harmonic wavelength of $286\,\mathrm{nm}$ and a power of $65\,\mathrm{mW}$. It is frequency stabilized using an ultra-low-expansion (ULE) material cavity. The laser light is relatively weakly focused and aligned parallel to the optical trap axis such that the intensity is effectively uniform over the atom cloud. Fig.~\ref{fig:ExperimentandRamsey}c shows the corresponding Ramsey signal. In comparison to the case of the undressed Ramsey interferometer (Fig.~\ref{fig:ExperimentandRamsey}b), we observe an asymmetric decay of contrast and a visible shift of the oscillation frequency.
	
To interpret the effects of Rydberg dressing we derive a simple analytical formula for the Ramsey signal, including the influence of noise on the Rydberg dressing laser and neglecting interaction effects, which is justified given the low densities used for the experiment. We start by considering an ensemble of identical four-level atoms, comprised of the clock states $\ket{0}$,$\ket{1}$, the Rydberg state $\ket{r}$, and an auxiliary shelving state $\ket{s}$ which collectively describes all states into which the Rydberg states can decay and that no longer participate in the dynamics. The system evolution is well described by a quantum master equation for the density matrix $\rho$ in Lindblad form: $\dot \rho = -i[\hat H,\rho]\,+\,\mathcal L (\rho)$. Here $\hat H$ accounts for the coherent dynamics and $\mathcal L$ is a superoperator describing irreversible dissipative processes. For the moment we assume that the only dissipative process is decay from $\ket{r}$ to $\ket{s}$ state with rate $\Gamma$ and processes that bring population back out of the $\ket{s}$ state can be neglected. Accordingly, the quantum master equation can be reduced to a Schr\"odinger equation governed by the non-Hermitian Hamiltonian: 
\begin{equation}
\hat H_\mathrm{NH} = \frac{\Omega}{2}\left( \ket{1}\!\bra{r}+\ket{r}\!\bra{1}\right )-\left (\Delta+ \frac{i \Gamma}{2}\right )\ket{r}\!\bra{r} - \delta \ket{0}\!\bra{0},
\end{equation} 
where $\Omega,\Delta$ are the amplitude and detuning of the Rydberg laser coupling and $\delta$ is the detuning of the RF field from the clock transition. Rydberg state decay appears as an additional imaginary detuning $i\Gamma/2$ which describes overall population loss from the Rydberg state out of the system with rate $\Gamma$.

In the weak dressing limit $|\Delta|\gg \Omega,\Gamma$, the $\ket{r}$ state can be adiabatically eliminated yielding an effective Hamiltonian for the slowly evolving clock states,
\begin{equation}
\hat H_\mathrm{eff} =  f_r\left(\Delta-\frac{i \Gamma}{2}\right )\ket{1}\!\bra{1} - \delta \ket{0}\!\bra{0},
\end{equation} 
where $f_r=\Omega^2/(\Gamma^2+4\Delta^2)$ can be recognized as the steady state Rydberg fraction of the dressed $\ket{1}$ state.
Finally, intensity and frequency noise of the dressing laser can be included by treating the light shift as a fluctuating quantity $E_\mathrm{LS} \rightarrow E_\mathrm{LS} + \sqrt{X}\xi(t)$, where $\xi(t)$ is assumed to be a zero mean white noise process with $\langle\xi(t)\xi(t')\rangle=\delta(t-t')$. We define the noise variance, 
\begin{equation}\label{eq:var}
X \approx \left(\epsilon_I^2+\epsilon_\Delta^2\right)\frac{\Omega^4}{16\Delta^2},
\end{equation}
in terms of the spectral densities of the relative intensity and detuning noise [$\epsilon_I^2$ and $\epsilon_\Delta^2$ respectively, with units $\mathrm{(frequency)}^{-1}$].

Following~\cite{plankensteiner2016} we write the stochastic time-dependent von Neumann equation for the density matrix $\rho_\mathrm{eff}$ as,
\begin{equation}\label{eq:me}
\dot \rho_\mathrm{eff} = -i [H_\mathrm{eff},\rho_\mathrm{eff}]- i \sqrt{X} \xi(t)[\ket{1}\bra{1},\rho_\mathrm{eff}].
\end{equation}
Performing the Markov approximation and using the identity for multiplicative linear white noise in~\cite{plankensteiner2016}, we arrive at
\begin{equation}\label{eq:me1}
\dot\rho_\mathrm{eff} = \left( \mathcal L_0+\frac{1}{2}X\mathcal L_1^2\right )\rho_\mathrm{eff}(t),
\end{equation}
where $\mathcal L_0\rho_\mathrm{eff}=-i[H_\mathrm{eff},\rho_\mathrm{eff}]$ and $\mathcal L_1\rho_\mathrm{eff}=-i[\ket{1}\bra{1},\rho_\mathrm{eff}]$.

Considering the Ramsey sequence in Fig.~\ref{fig:ExperimentandRamsey}a and applying Eq.~\eqref{eq:me1} during the free evolution time, the population in the $\ket{1}$ state after the second $\pi/2$ pulse is:
\begin{equation}\label{eq:pop}
	P_1 = \frac{1}{4} \left( 1 + e^{-f_r \Gamma t} + 2 e^{-\frac{1}{2}(f_r \Gamma + X)t}\cos\left[(\delta+f_r\Delta)t+\phi\right] \right),
\end{equation}
where we have included the possibility of a phase shift $\phi$ caused by the detuning during the two $\pi/2$ pulses. The effects of Rydberg-dressing can be identified as a differential energy shift between the $\ket{0}$ and $\ket{1}$ states, which changes the oscillation frequency by an amount $E_\mathrm{LS}=f_r\Delta$, an overall population decay with rate $\Gamma_\mathrm{eff}=f_r\Gamma$ and pure dephasing rate $X$. 

\begin{figure}[t!]
	\includegraphics[width = 0.99\columnwidth]{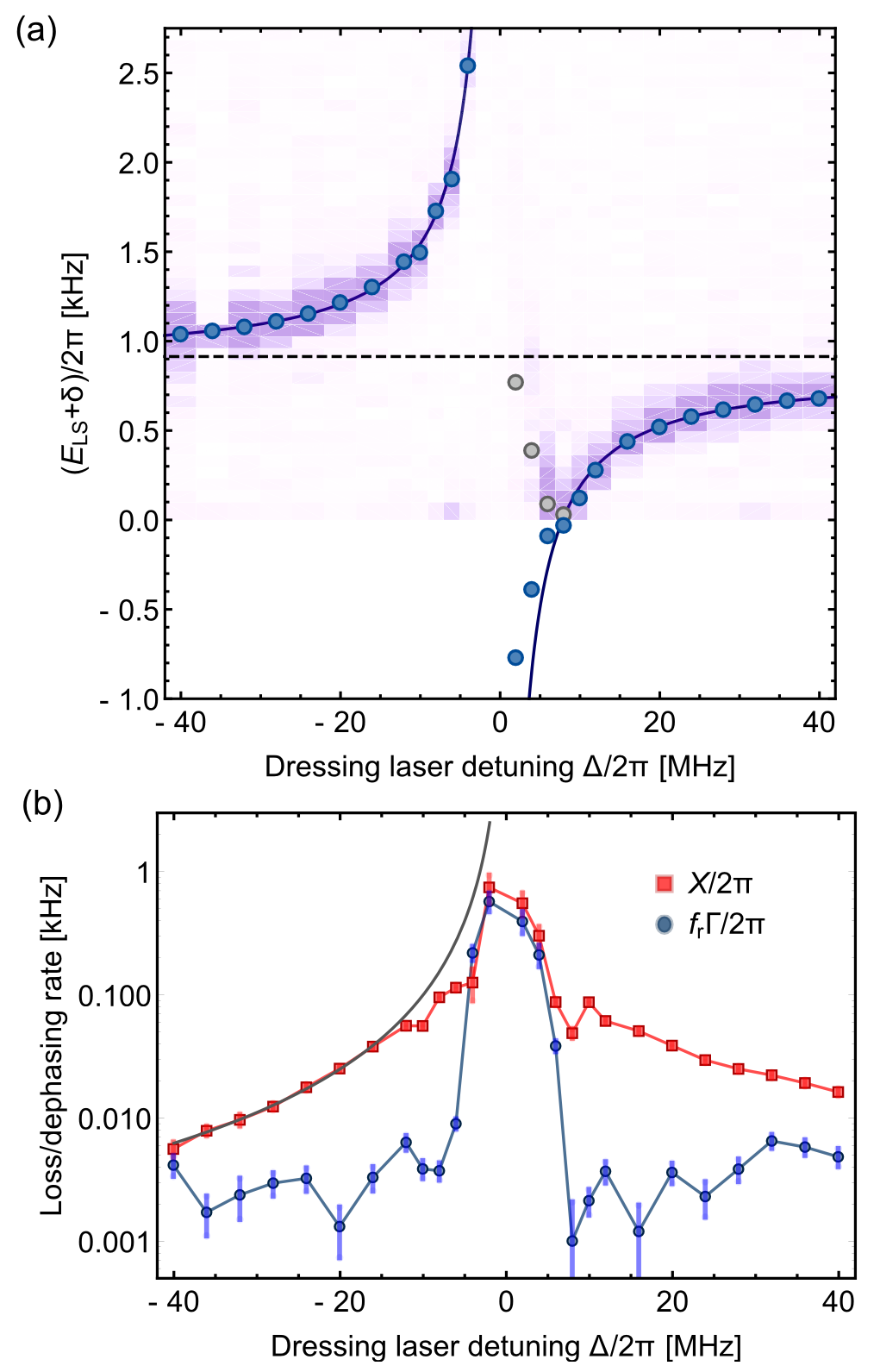}
	\caption{Analysis of the Ramsey signals as a function of the detuning of the Rydberg dressing laser $\Delta$. (a) Ramsey oscillation frequency (blue circles) in the presence of the dressing laser extracted from fits to Eq.~\ref{eq:pop}. Gray circles correspond to positive oscillation frequencies before mapping the data. The statistical errors for each measurement are $\lesssim50\,$Hz (well within the size of the data points). The blue line is a fit to the theoretical model for the dressed-state energy used to determine the atom-light coupling strength $\Omega$. The dashed line indicates the detuning $\delta$ of the bare clock states. The shaded background shows the Fourier transform of each Ramsey signal.  (b) Dressed-state dephasing rate (red squares) and population decay rate (blue circles) as a function of $\Delta$. The black curve shows the power law $\Delta^{-2}$ which is the expected dependence for intensity-noise dominated dephasing.}
	\label{fig:DephasingAndLoss}
\end{figure}

In the following we analyze the performance of the Rydberg-dressed Ramsey interferometer for different detunings of the dressing laser~$\Delta$. For this, we fit the Ramsey measurements to Eq.~\eqref{eq:pop} including an experimentally determined phaseshift $\phi=0.55$. As exemplified by the solid lines in Fig.~\ref{fig:ExperimentandRamsey}b,c the model fits the data extremely well such that we can extract the detuning dependent parameters $E_\mathrm{LS}$, $\Gamma_\mathrm{eff}$, and $X$. 

\noindent\emph{Light shifts:--} Figure~\ref{fig:DephasingAndLoss}a shows the Rydberg induced light shift $E_\mathrm{LS}$ as read out from the oscillation frequency for different detunings of the Rydberg dressing laser in the range $\pm 40$~MHz. The density plot in the background of the figure corresponds to the Fourier transform of each respective Ramsey signal. To account for the cases close to resonance where the oscillation frequency $E_\mathrm{LS}+\delta$ becomes negative we map the sign of the data points such that the frequency is monotonic as a function of $\Delta$ (these data points before mapping are displayed in gray in Fig.~\ref{fig:DephasingAndLoss}a). 

Although the measured light shifts are relatively small, $\lesssim 1\,\mathrm{kHz}$ over most of the range, they are clearly resolved with an average statistical uncertainty of $\sim2\times 10^{-3}$ for $|\Delta|/2\pi>10\,\mathrm{MHz}$. Within this accuracy we observe small deviations from the simple two-level weak-dressing prediction of $E_\mathrm{LS}=\Delta\Omega^2/(\Gamma^2+4\Delta^2)$ due to the nearby transition $\ket{0}\rightarrow \ket{39P_{1/2}}$ which is resonant for $\Delta/2\pi=76\,\mathrm{MHz}$. The quality of the fit that takes into account the differential light shift including the extra state, shown as a solid line in Fig.~\ref{fig:DephasingAndLoss}a, reflects a very good quantitative understanding of the dominant effects influencing the frequency of the Rydberg-dressed Ramsey interferometer. It also confirms that interaction effects such as the Rydberg blockade do not play an important role here. 

From the fit shown in Fig.~\ref{fig:DephasingAndLoss}a we obtain a precise determination of the Rabi frequency $\Omega/2\pi=163(1)\,\mathrm{kHz}$. This value is more than a factor of two smaller than an independent estimate based on the power and waist of the laser beam and the expected transition dipole matrix element. This highlights the importance of direct experimental measurement of the atom-light interaction parameters over indirect estimates. The range of resolved light shifts corresponds to Rydberg fractions from approximately $4 \times 10^{-4}$ to a minimum value of $4 \times 10^{-6}$, corresponding to less than one Rydberg excitation shared amongst the entire cloud, emphasizing the sensitivity of the Ramsey interferometer to very small perturbations.

\noindent\emph{Contrast:--} The decaying contrast of the Ramsey fringes gives further information on the coherence of Rydberg-dressed states which limits the interferometer performance. While dephasing due to laser intensity or frequency noise should result in a symmetric loss of contrast with respect to the mean, population loss causes an overall reduction of the mean value. As such, these contributions can be separately extracted from the fits to Eq.~\eqref{eq:pop} and are shown in Fig.~\ref{fig:DephasingAndLoss}b on a logarithmic scale. Close to resonance we find that the contributions are approximately equal with a rate around $1\,\mathrm{kHz}$. This is comparable to the bare Rydberg state decay rate estimated as $\Gamma/2\pi=5.6\, \mathrm{kHz}$ which includes photoionization from the optical dipole trap, blackbody redistribution and spontaneous decay. The maximum dephasing rate is two orders of magnitude smaller than the independently measured laser linewidth showing that the correspondence between laser phase noise and the loss of Ramsey coherence can be subtle.

For large detunings $|\Delta|/2\pi>5\,\mathrm{MHz}$ we find that dephasing is the dominant effect governing the loss of coherence rather than intrinsic atom loss. Nevertheless the combined loss and dephasing rates fall below $0.1\,$kHz corresponding to a usable Ramsey evolution time of tens of milliseconds, which is orders of magnitude longer than the bare Rydberg state lifetime and sufficiently long to allow precision measurements. This is demonstrated by a large ratio of the induced light shift to the dephasing rate, quantified by the strong-coupling parameter $C=E_\mathrm{LS}/(\Gamma_\mathrm{eff}+X)$, that varies in the range $7\leq C\leq18$ as $\Delta$ is varied over the range $\pm 40\,\mathrm{MHz}$. Looking closer at the detuning dependent dephasing rate we find that it is well described by a power-law, which below resonance, scales with $\Delta^{-1.98(4)}$. This is close to the expected scaling for intensity noise, which according to Eq.~\eqref{eq:var} should scale as $X\propto |\Delta|^{-2}$, as opposed to laser frequency noise which should scale with $|\Delta|^{-4}$ since the relative frequency noise $\epsilon_{\Delta}^2$ scales as $|\Delta|^{-2}$. From the power-law fit we extract the normalized intensity spectral density: $2\pi\epsilon_I^2=0.2\,\mathrm{kHz}^{-1}$ which is compatible with residual intensity noise. 

\begin{figure}[t!]
	\includegraphics[width = 0.99\columnwidth]{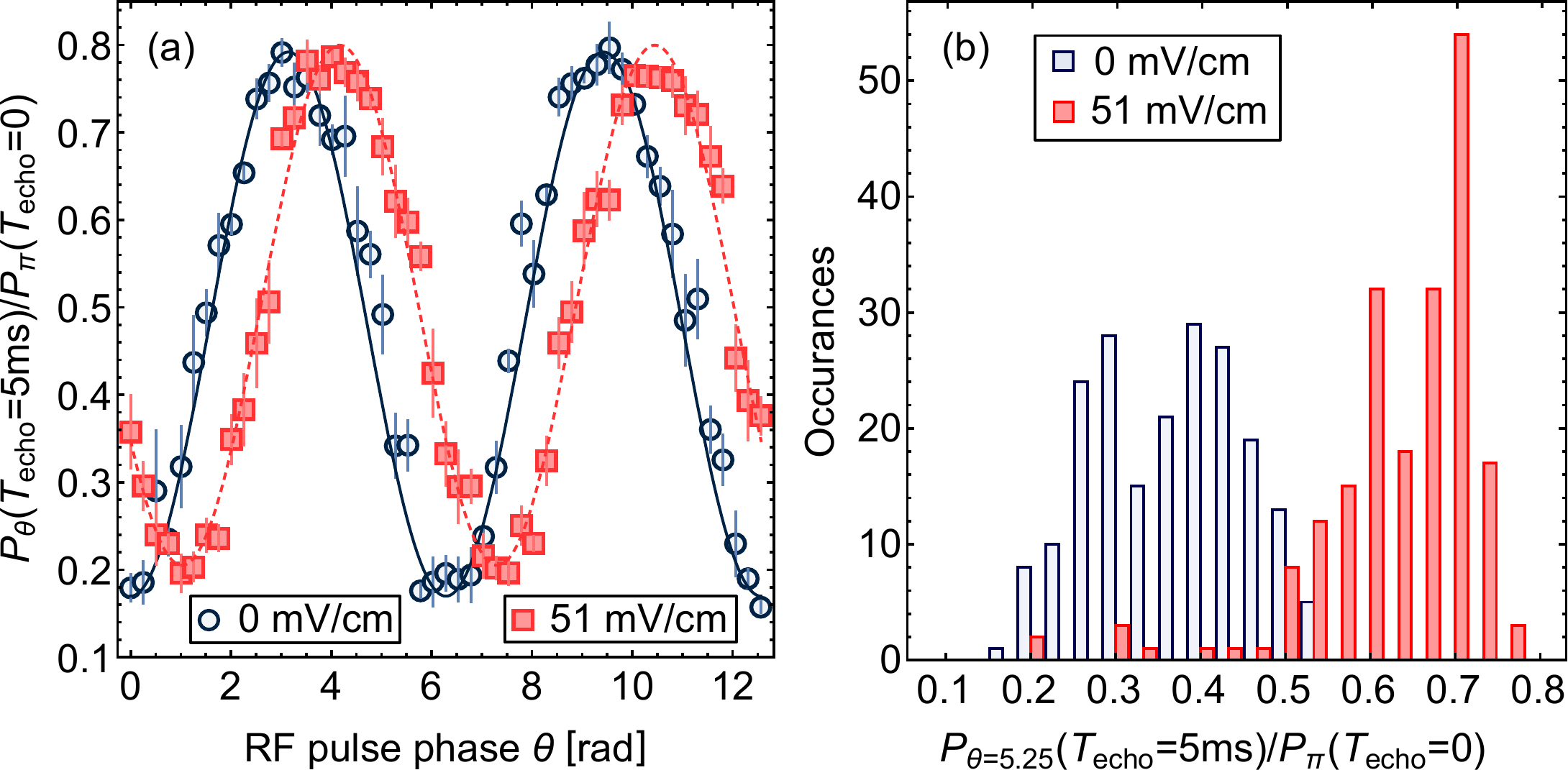}
	\caption{Sensitivity of the Rydberg dressed interferometer to small electric fields. (a) Interferogram obtained via spin echo measurement for two different perturbing electric fields for a free evolution time of $T_{\mathrm{echo}}=5\,$ms, and with a bias electric field of $F=2.2\,\mathrm{V / cm}$.  The normalization of the population is for $T_{\mathrm{echo}}=0\,$ms, with $\theta = \pi$, which is when the population is maximum.  (b) Histograms of 200 measurements at $\theta = 5.25$\,rad for the two electric field values.}
	\label{fig:Electrometer}
\end{figure}

Having characterized the performance of the Rydberg-dressed interferometer, we now demonstrate its enhanced sensitivity to static electric fields via the induced polarizability of the Rydberg-dressed state (Fig.~\ref{fig:Electrometer}). For this we use a spin echo RF pulse sequence consisting of two $\pi/2$-pulses separated by two free evolution times of equal duration $T_{\mathrm{echo}}=5\,$ms, with a $\pi$-pulse in between. The UV dressing laser is applied continuously during both of the free evolution times while an applied electric field is switched from $F_1=2.2\,\mathrm{V / cm}$ to $F_2=F_1+\epsilon$ between the two free evolution times. In this way common mode effects largely cancel out, e.g. energy shifts caused by the bias electric field and dressing laser light shifts. On the other hand, the electric field perturbation $\epsilon$ causes a phase shift of the Ramsey interferograms obtained as a function of the phase $\theta$ of the final $\pi/2$-pulse. Exemplary results are shown in Fig.~\ref{fig:Electrometer}(a) for $\epsilon=0$ (open circles) and $\epsilon=51(2)\,\mathrm{mV / cm}$ (red squares) clearly showing an electric field dependent phase shift corresponding to $0.022(1)\,$rad/(mV/cm). To determine the single shot sensitivity of our setup we repeat the measurement 200 times for each electric field setting with the Ramsey phase set to a constant value of $\theta = 5.25$~rad. The resulting statistical distributions shown in Fig.~\ref{fig:Electrometer}(b) are clearly separated, by three standard deviations, implying a single shot sensitivity of 17(1)~$\mathrm{mV / cm}$.

To calculate the fundamental sensitivity limit we maximize the derivative of Eq.~\ref{eq:me} with respect to small perturbations. Assuming shot noise limited state readout we find a minimum single shot sensitivity (standard quantum limit) $\sigma=e (X+f_r\Gamma)/(f_r F \alpha \sqrt{N})$, where $e$ is Euler's number, $N$ is the number of atoms used in the measurement and $\alpha$ is the electric polarizability of the bare Rydberg state (which scales with principal quantum number as $n^7$). Importantly, the limit obtained for $X\rightarrow 0$ is independent of the Rydberg state fraction. Thus, the Rydberg dressing approach allows for extremely small densities of the strongly interacting particles, while at the same time enabling large overall atom numbers to minimize the atomic shot noise. For the Rydberg state used in our experiment (having $\Gamma/2\pi=5.6\,\mathrm{kHz}$ and $\alpha_{39P}/2\pi=19.3\,\mathrm{MHz\,cm^2 / V^2}$) with $N=10^5$ particles and applying a bias field $F=2\,\mathrm{V/cm}$, we find $\sigma=1.2\,\mathrm{\mu V/cm}$ which indicates that such Rydberg-dressed electrometers would rival the state-of-the-art electrometers~\cite{cleland1998nanometre,Bunch2007,facon2016sensitive,kumar2017atom}.

While the present experiments were performed in an effectively non-interacting regime, Rydberg-dressed Ramsey interferometry can serve as a powerful method to induce, control and characterize atomic interactions~\cite{Nipper2012,Martin2013,Hazzard2014,Ebert2015,takei2016,zeiher2016many}. This could provide a tunable, nonlinear evolution between the clock states for generating squeezed or entangled many-body states without relying on low-energy collisions between the atoms~\cite{Bouchoule2002,riedel2010,gross2010,Gil2014}, or a realization of nonlinear quantum metrology protocols~\cite{Luis2004,Boixo2008} for surpassing the standard quantum limit. Alternatively, Ramsey interferometers provide a way to optimize Rydberg-dressing protocols~\cite{Balewski2014,Jau2015,zeiher2016many,deleseleuc2018}, even in the limit of extremely weak interactions, helping to realize robust quantum logic gates and quantum spin systems~\cite{Keating2015,bernien2017probing,Zeiher2017coherent,picken2019Entanglement}, or novel long-range interacting quantum fluids~\cite{Wang2006,Dalmonte2010} and lattice gases with beyond nearest neighbor interactions~\cite{Yan2013a,DePaz2013,baier2016extended}.  

In conclusion, we have realized and characterized the performance of a trapped atom Ramsey interferometer utilizing clock states enhanced by Rydberg dressing. This includes a precise determination of the atom-light coupling, Rydberg state admixture and coherence properties of the dressed states. We also demonstrated an application for measuring small electric field perturbations. The basic essence of our scheme -- the controllable coupling of a highly-coherent two-level system to a third state with greatly enhanced sensitivity -- can be applied to numerous quantum systems, even beyond the Rydberg atom platform. Thus this approach has the potential to greatly expand the number of atomic and molecular systems suitable for metrological applications and many-body physics.

\acknowledgements{We acknowledge early contributions to the experiment by H. Hirzler as well as valuable discussions with J. Schachenmayer. This work is part of and supported by the DFG Collaborative Research Centre ``SFB 1225 (ISOQUANT)'',  the Heidelberg Center for Quantum Dynamics and the ``Investissements d'Avenir'' programme through the Excellence Initiative of the University of Strasbourg (IdEx). S.W. was partially supported by the University of Strasbourg Institute for Advanced Study (USIAS), A.A. and S.H. acknowledge support by the Heidelberg Graduate School for Fundamental Physics, S.H. acknowledges also support by the Carl-Zeiss foundation.}

\bibliography{library}

\end{document}